\documentclass[sigconf]{acmart}
\usepackage{graphicx}
\usepackage{subcaption}
\usepackage{pifont}
\usepackage{enumitem}
\usepackage{makecell}
\usepackage[nameinlink]{cleveref}
\usepackage{multirow}
\usepackage{algorithm2e}
\usepackage{framed}
\usepackage{xcolor}
\usepackage{colortbl}

{\newcommand{\mynote}[2]{
\fbox{\bfseries\sffamily\scriptsize#1}
{\small$\blacktriangleright$\textsf{\emph{#2}}$\blacktriangleleft$}}}
{ \newcommand{\mynote}[2]{}}

{\endMakeFramed}

\newcommand{\find}[1]{%
  \par\noindent%
  \colorbox{gray!15}{%
    \parbox{\dimexpr\linewidth-2\fboxsep}{
      \em #1
    }%
  }%
  \par
}

\definecolor{mylightgray}{RGB}{224,224,224}

\newcommand{\wcircle}[1]{\ding{\numexpr171 + #1}}
\newcommand{\bcircle}[1]{\ding{\numexpr181 + #1}}

\copyrightyear{2026}
\acmYear{2026}
\setcopyright{rightsretained}
\acmConference[ICSE '26]{2026 IEEE/ACM 48th International Conference on Software Engineering}{April 12--18, 2026}{Rio de Janeiro, Brazil}
\acmBooktitle{2026 IEEE/ACM 48th International Conference on Software Engineering (ICSE '26), April 12--18, 2026, Rio de Janeiro, Brazil}
\acmPrice{}
\acmDOI{10.1145/3744916.3764554}
\acmISBN{979-8-4007-2025-3/26/04}

\ccsdesc[100]{General and reference~Empirical studies}
\ccsdesc[500]{General and reference~Evaluation}
\ccsdesc[500]{Computing methodologies~Natural language generation}
\ccsdesc[300]{General and reference~Performance}

\author{Djiré Albérick Euraste}
\affiliation{
  \institution{University of Luxembourg \&}
  \institution{AI4D Excellence Centre (CITADEL)}
  \city{Ouagadougou}
  \country{Burkina Faso}
}

\author{Kaboré Abdoul Kader}
\affiliation{
  \institution{University of Luxembourg}
  \city{Luxembourg}
  \country{Luxembourg}
}

\author{Jordan Samhi}
\affiliation{
  \institution{University of Luxembourg}
  \city{Luxembourg}
  \country{Luxembourg}
}

\author{Earl T. Barr}
\affiliation{
  \institution{University College London}
  \city{London}
  \country{UK}
}

\author{Jacques Klein}
\affiliation{
  \institution{University of Luxembourg}
  \city{Luxembourg}
  \country{Luxembourg}
}

\author{Tegawendé F. Bissyandé}
\affiliation{
  \institution{University of Luxembourg}
  \city{Luxembourg}
  \country{Luxembourg}
}

\renewcommand{\shortauthors}{Djiré, et al.}

\title[Quantifying Memorization Advantage in Code LLMs]{Learned or Memorized ? Quantifying Memorization Advantage in Code LLMs}
\begin{document}

\begin{abstract}
The lack of transparency regarding the code datasets used during LLM training creates substantial challenges in detecting, evaluating, and mitigating data leakage. This paper applies a perturbation-based approach to quantify the ``memorization advantage'' of LLMs across various coding tasks by measuring the performance gap between a model's handling of data it has likely encountered during training versus novel inputs. Our comprehensive analysis examines 8 open-source code LLMs across 19 benchmark datasets spanning four distinct categories: standard code generation, code understanding, security vulnerability detection, and bug identification. The results reveal significant variations in sensitivity patterns, with models like StarCoder exhibiting substantially higher sensitivity scores (up to 0.8) on certain benchmarks like APPS compared to models like QwenCoder, which maintained consistently lower values ($<$0.4) across most benchmarks, suggesting fundamental differences in their generalization process and their learned knowledge. Different task categories also showed distinct patterns, with code summarization demonstrating low sensitivity ($<$0.3) and test generation tasks exhibiting significantly higher values (0.4-0.7, $p<0.001$).

Interestingly, our analysis of the widely-used CVEFixes and Defects4J benchmarks, frequently suspected of data leakage in the research community, reveals unexpectedly low memorization advantage scores across all models. Defects4J demonstrated significantly lower sensitivity (0.2-0.4, $p<0.01$) compared to other program repair benchmarks (0.5-0.8), while CVEFixes showed consistently low values below 0.1. These findings challenge prevailing concerns about these datasets' validity for evaluating code LLMs and suggest that models may be effectively generalizing from this data rather than merely memorizing it. Our findings provide critical insights into the generalization capabilities of code LLMs and emphasize the need for more robust evaluation frameworks, particularly in security-related domains where the widest range of sensitivity distributions (from $<$0.1 to $>$0.8) indicates variable generalization challenges.
\end{abstract}

\maketitle

\section{Introduction} \label{sec:introduction}

Large Language Models (LLMs) represent a paradigm shift in computational linguistics, fundamentally transforming how systems process and generate human language in various tasks such as text completion or translation. However, the capabilities of these models now extend beyond traditional linguistic boundaries into other fields such as software engineering where the ``naturalness''~\cite{hindle2016naturalness} of code constitutes fertile ground for the application of LLMs.

Recent advances in LLM reasoning capabilities have significantly improved their problem-solving ability, allowing them to tackle increasingly complex computational challenges through structured analytical processes~\cite{guo2025deepseek}. This has directly translated to better performance in code-related tasks, where models now demonstrate advanced abilities in algorithmic design or prediction of logical sequences. Today, a large family of specialized models targeting code analysis and processing have been presented in the literature, achieving state-of-the-art performance in a variety of software-related tasks, including test generation \cite{xie2023chatunitest, zhang2023well}, vulnerability detection \cite{chakraborty2020deeplearningbasedvulnerability, risse2024uncovering}, code summarization \cite{sontakke2022code}, and program repair \cite{deligiannis2023fixing, fu2022vulrepair, jin2023inferfix}. 

However, the promising reported results come with recurrent concerns about their reliability, notably due to suspicions of data leakage~\cite{zhou_dont_2023, balloccu_leak_2024, matton_leakage_2024}. Indeed, because of privacy or commercial considerations, the literature often omits key information on the datasets that were used to train or fine-tune models. Yet, in the absence of data transparency, it becomes difficult to assess the true generalization capabilities of an LLM for a given task, since the claimed performance improvements might stem from prior exposure to evaluation data during training, leading to inflated results.

Measuring data leakage is very challenging. Many existing approaches require knowledge either about the internal details of the models~\cite{carlini_quantifying_2023, dankers_generalisation_2024} or some details of part of datasets used during training~\cite{carlini_extracting_nodate}. More recently, novel ideas have emerged that reason about the observations of how LLMs outputs behave when inputs are perturbed \cite{xie_memorization_2024}. In this paper, we build on these ideas and implement an approach that evaluates the sensitivity of the LLM to subtle input perturbations in order to measure its \textbf{memorization advantage}. This advantage is defined as \textit{the difference in performance between when the model is dealing with 'seen' versus when it is dealing with 'unseen' data}. Formally, the memorization advantage $ma$ for a model $M$ can be defined as:
\begin{center}
$ ma(M, x, y) = | p_\theta(y|x) - p_\theta(y|x'(x)) | $
\end{center}
Where $(x, y)$ is a sequence from the training data,  $x'(x)$ is a function that returns a similar but unseen sequences and $p_\theta(y|x)$ is the model's probability of generating $y$ given $x$.

{\bf This paper.} Given persistent debates in the community around the likely inflated performance of LLMs due to potential data leakage, we propose an extensive investigation to shed light into the extent of the problem. Our empirical experiments consider eight (8) popular code LLMs which we apply to twenty (20) different benchmarks across five (5) classical tasks of software engineering: code summarization, vulnerability detection, test generation, program repair, and  code synthesis.  

We examine the phenomenon (memorization advantage) from two complementary viewpoints: one focusing on identifying potentially contaminated benchmarks, and the other on identifying models with unusual performance patterns on certain data:

\begin{itemize}[leftmargin=*]
    \item \textbf{Benchmark Memorization Advantage}: Given a model $m$ and a set of benchmarks, $B = \{b_0, b_1, ... b_n\}$, we quantify the memorization advantage of a subset \(B^* \subset B\) as a statistically significant difference in distribution of input sensitivity scores on this subset compared to the rest, suggesting potential data contamination or memorization of \(B^*\) by $m$.

    \item \textbf{Model Memorization Advantage}: Given a benchmark $b$ and a set of models, $M = \{m_0, m_1, ..., m_n\}$, we quantify the memorization advantage of a model \(m_i\) as the statistically significant difference in the distribution of input sensitivity scores of \(m_i\) compared to others on $b$, suggesting that \(m_i\) has a systematic advantage in the evaluated data set.

\end{itemize}

The main contributions of this paper are:

\noindent
\bcircle{1}
Application of prompt perturbation sensitivity analysis to quantify memorization advantage in code LLMs, providing an empirical framework for distinguishing between memorization and generalization effects in model performance
    
\noindent
\bcircle{2}
Empirical evaluation of 8 state-of-the-art code LLMs across 19 benchmark datasets spanning code generation, test generation, program repair, and vulnerability detection tasks.
        
\noindent
\bcircle{3}
Statistical analysis results challenging prevailing assumptions about data leakage in widely-used benchmarks (particularly CVEFixes and Defects4J), revealing unexpectedly low memorization advantage scores that suggest genuine generalization rather than memorization
    
\noindent
\bcircle{4}
Identification of significant variations in sensitivity patterns across models with comparable parameter counts, demonstrating that architectural design choices and training methodologies substantially impact generalization capabilities
    
\noindent
\bcircle{5}
Task-specific insights showing that code summarization tasks exhibit the strongest generalization across all models, while test generation and context-dependent program repair present the greatest generalization challenges

\noindent
\textbf{Artifacts.}
We make all our artifacts available at:
\begin{center}
    \url{https://github.com/Berickal/CodeLLM_Memo}
\end{center}

\section{Background} 
\label{sec:background}
\subsection{Memorization vs. Generalization in Language Models}

LLMs function as probabilistic systems that extract and generalize patterns from training data. The fundamental challenge these models face is balancing between pattern generalization and maintaining fidelity to their training data for reliable output generation~\cite{carlini_extracting_nodate, hartmann2023sok}. This creates an inherent tension in model development and evaluation.

The literature presents multiple definitions for memorization in LLMs. A prominent definition focuses on exact sequence reproduction, where models output verbatim content from their training data~\cite{carlini_extracting_nodate, ozdayi_controlling_2023}. This phenomenon appears across various tasks including completion, question-answering, and logical reasoning~\cite{hartmann2023sok}. Though memorization is typically associated with overfitting during fine-tuning phases~\cite{schwarzschild2024rethinking, duan2024membership, speicher_understanding_2024}, recent research suggests it may actually be a prerequisite for effective generalization, with models showing higher sensitivity to memorized outliers than to common patterns~\cite{dankers2024generalisation}.

Generalization, by contrast, describes a model's capability to identify and leverage underlying patterns from training data to generate novel outputs. Formally, for an input $x$ from space $\mathcal{X}$, a generalizing model produces outputs by computing the expected value over its learned distribution:

\begin{equation}
f(x) = \mathbb{E}_{y \sim p(y|x)}[y]
\end{equation}
where $f(x)$ represents the model's output, and the expected value is calculated over outputs $y$ sampled from the conditional probability distribution $p(y|x)$ that the model has learned.

\subsection{Challenges in Detecting Memorization in Code LLMs}

Detecting memorization in code LLMs presents unique challenges compared to general-purpose language models:

\subsubsection{Data Provenance Opacity}

Unlike some general-purpose LLMs with partially documented training sources, code LLMs typically lack transparency regarding training datasets~\cite{bommasani2022opportunities}. This opacity complicates determination of which code examples a model has encountered during training.

\subsubsection{Code Similarity Evaluation}

Code memorization detection is complicated by implementation variability for similar functionality. Standard text similarity metrics may inadequately determine whether a model has memorized specific implementations or genuinely understood programming concepts~\cite{allamanis2018survey, peng2021could}.

\subsubsection{Benchmark Contamination}

Many standard code benchmarks are publicly available and may be included in various code LLMs' training data~\cite{sharma2022evaluation}. This potential contamination undermines traditional evaluation approaches for assessing true model capabilities.

\subsection{Perturbation-Based Approaches for  LLM Robustness Evaluation}

Perturbation-based approaches offer promising methodologies for evaluating LLMs by examining performance changes when inputs undergo systematic modification. Literature examples include:

\begin{enumerate}[leftmargin=*]
    \item \textbf{Adversarial Perturbations}: Research has demonstrated how character-level and word-level perturbations reveal vulnerabilities in language models~\cite{ebrahimi2018hotflip, jin2020bert}, with specific extensions to code models~\cite{wallace2021analyzing}.
    
    \item \textbf{Counterfactual Data Augmentation}: Techniques generating counterfactual examples through minimal input perturbations reveal model robustness characteristics~\cite{kaushik2020learning, gardner2020evaluating}.
    
    \item \textbf{Program Transformation Analysis}: Studies have explored how systematic code transformations affect model performance, providing insights into code model robustness~\cite{rabin2021understanding, hartmann2023sok}.
\end{enumerate}

For code LLMs, various types of perturbations can be considered: (1) \textbf{Syntactic Perturbations}: Modifications preserving program semantics while altering syntax, such as variable renaming or code reformatting~\cite{rabin2021understanding, jha2022codeattack}. (2) \textbf{Semantic Perturbations}: Changes slightly altering program behavior, like modifying constant values or changing loop boundaries~\cite{gupta2020synthesize}. And (3) \textbf{Structural Perturbations}: Alterations to code organization while preserving functionality, including function reordering or module restructuring~\cite{wang2020detecting}.

\subsection{The Perturbation Sensitivity Hypothesis\newline on Memorization}

\begin{figure}[!ht]
     \centering
     \vspace{-0.45cm}
     \includegraphics[width=1\linewidth]{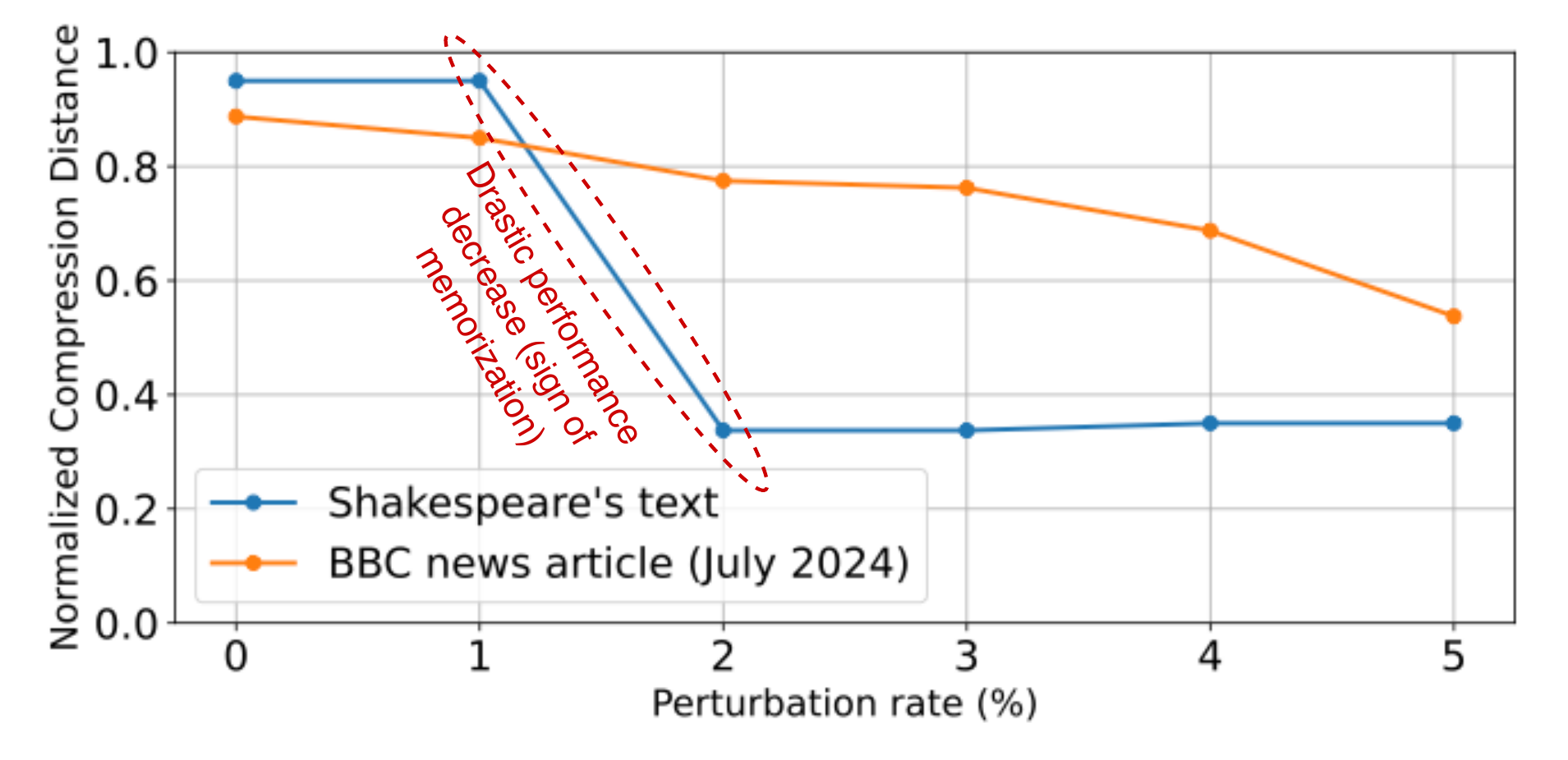}
     \caption{GPT\_4o text completion \textit{performance falloff} for a memorized Shakespeare poem submitted to perturbations vs gradual performance decline with a recent BBC text (could not part of the training set of GPT\_4o).}
     \label{fig:motivation_example}
     \Description{Performance evolution of a memorized Shakespeare poem under bit-flip perturbation was compared to the gradual degradation of a recent text not included in GPT-4’s training set. The Shakespeare poem showed a significant variation at a 2\% perturbation rate, highlighting the model's sensitivity to memorized content, in contrast to the latter text.}
 \end{figure}

Prior work has demonstrated that neural network generalization capabilities can be evaluated through sensitivity to data perturbations~\cite{zhang2021understanding, cohen-inger_forget_2025}. When presented with text samples for completion tasks, models may produce correct outputs regardless of whether they memorized or interpolated. However, introducing perturbations reveals distinctive behavioral patterns. With memorized content, even minor perturbations (e.g., random bit flips at 2\% rate) cause abrupt performance degradation, while interpolated understanding shows more gradual performance decline as perturbation increases as illustrated in Figure~\ref{fig:motivation_example}.

This Perturbation Sensitivity Hypothesis (PSH) offers a methodological framework to distinguish between memorization and generalization in LLMs. According to PSH, models that have memorized specific content demonstrate high sensitivity to input perturbations, while models effectively interpolating show greater robustness when inputs are perturbed~\cite{feldman2020neural, cohen-inger_forget_2025}.

\subsection{Memorization Advantage in LLM Evaluation}

In this work, our definition of\textit{ memorization advantage} extends the perturbation sensitivity concept by quantifying performance disparities between a model's handling of familiar versus novel inputs. This metric is particularly valuable for evaluating LLMs when data leakage might significantly impact assessment results.

\subsubsection{Definition and Measurement}

Memorization advantage directly quantifies a model's sensitivity to input perturbations, reflecting the differential robustness between training and non-training data.

Given a task $T$, an input $X$ and the set of generated perturbed inputs \(X^* = \{x^*_0, x^*_1, ... x^*_n\}\), we prompt the LLM and collect a set of output sets \(Y^* = \{Y^*_0, Y^*_1, ... Y^*_n\}\) with \(Y^*_k = \{y^*_{(k, 1)}, y^*_{(k, 2)}, …. y^*_{(k, i)} \}\) being the set of outputs associated with the perturbed input \(x^*_k\). 
Indeed, for each input, we prompt the LLM $i$ times to obtain $i$ sample outputs. This is done in order to help establish statistical significance of results as single runs might be outliers and not representative of true model capabilities. 

For a given output set \(Y^*_k\) (yielded by an LLM prompted with perturbed input \(x^*_k\)) and a reference output \(Y\), we compute a metric on the performance variation as follows:
\begin{equation} \label{eq:2}
    m(Y^*_k) = \frac{1}{i} \sum_{j=1}^idistance(y^*_{(k,j)}, Y) 
\end{equation}
where $distance$ implements a task-dependent edit distance function.  

To quantify the model's sensitivity to the perturbations on input $X$, we compute the maximum performance falloff when the model is subjected to increasingly intense perturbations:
\begin{equation} \label{eq:3}
    sensitivity(X) = {\underset{j \in {1,..,k-1}}{max}(m(Y^*_j) - m(Y^*_{j+1})) }
\end{equation}

\subsubsection{Significance in Code LLMs}

For code-generating LLMs, memorization advantage carries particular importance due to several factors:

\begin{enumerate}[leftmargin=*]
    \item \textbf{Benchmark Contamination}: Popular code benchmarks may have been incorporated into training data, artificially inflating performance metrics~\cite{kandpal2022deduplicating, zhou2023large}.
    
    \item \textbf{Security Implications}: In security-critical applications like vulnerability detection, memorizing known vulnerabilities without genuine understanding could create false confidence in model capabilities~\cite{pearce2022examining}.
    
    \item \textbf{Generalization Assessment}: Differentiating between memorization and true generalization is essential for understanding a model's practical utility in software development contexts~\cite{chen2021evaluating, austin2021program}.
\end{enumerate}
\section{Experimental Setup} \label{sec:methodology}
This section presents our experimental framework for quantifying memorization advantage in code LLMs. We first describe our methodology for measuring perturbation sensitivity (Section~\ref{subsec:methodology}), followed by details on the evaluated models (Section~\ref{subsec:models}), the tasks and datasets used in our evaluation (Section~\ref{subsec:datasets}), and our perturbation techniques (Section~\ref{subsec:perturbations}). Together, these components form a robust evaluation pipeline that enables systematic analysis of memorization versus generalization across diverse code tasks.

\subsection{Methodology} \label{subsec:methodology}
Our methodology is grounded in the Perturbation Sensitivity Hypothesis (PSH), which posits that models demonstrate higher sensitivity to perturbations on memorized inputs compared to those where true generalization has occurred.

The core intuition driving our approach is that when a model has memorized specific code snippets or patterns, its performance will degrade significantly when those inputs undergo even minor perturbations. Conversely, when a model has successfully learned the underlying patterns and principles (interpolation), it will maintain more consistent performance despite input variations.

\begin{figure}[!h]
    \centering
    \includegraphics[width=1\linewidth]{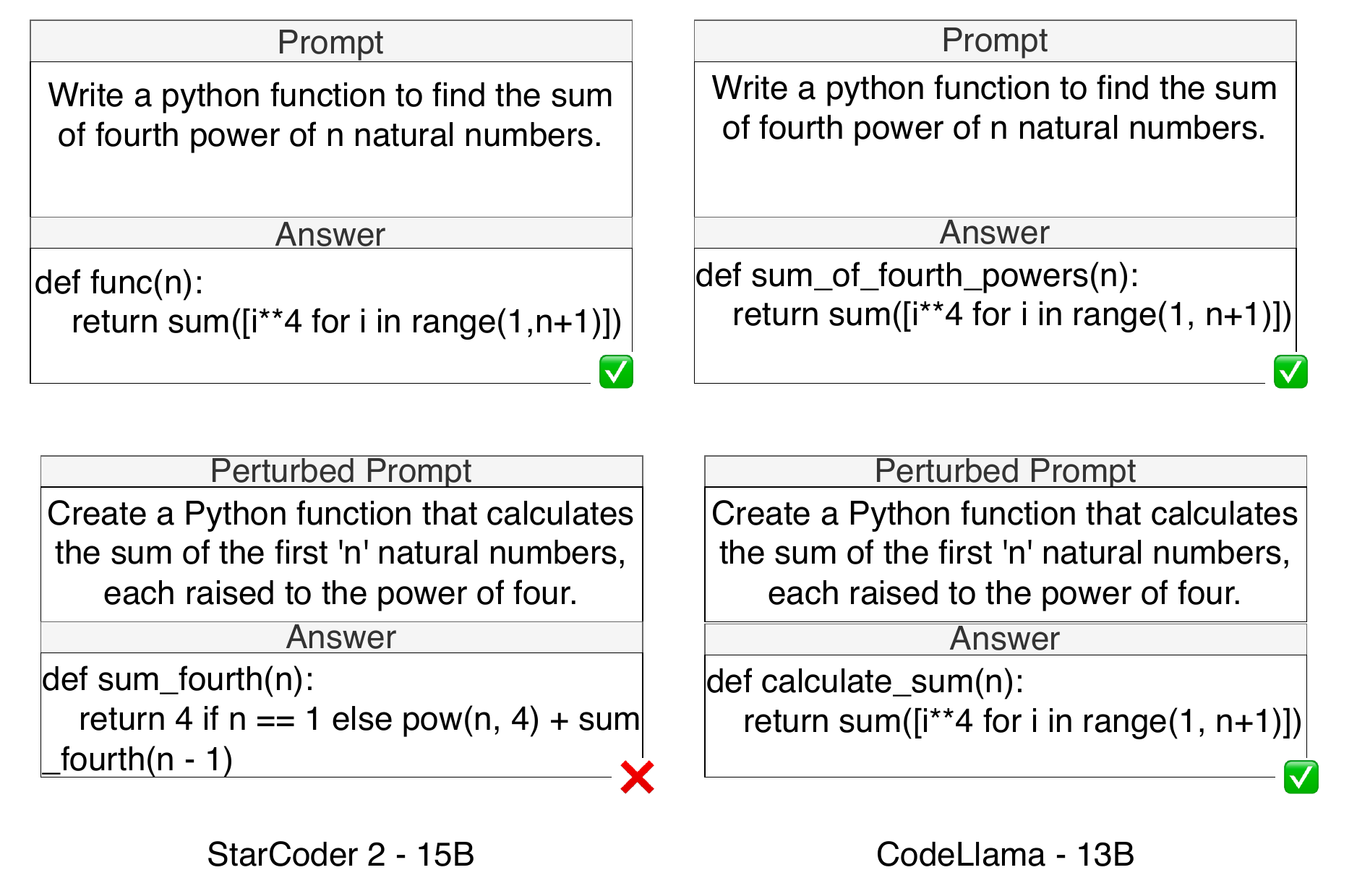}
    \caption{Memorization advantage example: impact of prompt perturbation on the outputs of StarCoder and CodeLlama}
    \label{fig:example}
    \Description{Memorization effects demonstrated through MBPP problem rephrasing. Both StarCoder2-15B and CodeLlama-13B successfully solved the original MBPP problem. However, when the problem statement was rephrased, only CodeLlama maintained its performance, revealing StarCoder2-15B's heavy reliance on memorized training data rather than genuine problem understanding.}
\end{figure}

Figure~\ref{fig:example} illustrates this principle with a comparison between StarCoder2 and CodeLlama on a code completion task on an example selected from the MBPP benchmark. The yielded outputs demonstrate how StarCoder2 exhibits higher sensitivity to "prompt" perturbations compared to CodeLlama, suggesting higher memorization advantage on this sample.

Algorithm~\ref{alg:pearl} formalizes our perturbation sensitivity evaluation procedure. For each input, \ding{182} we apply progressively stronger (but still small\footnote{Here, by small, we mean perturbations that would not change a human's response to the perturbed query.  Like adversarial examples, our perturbations are constrained to be those a human would ignore or fail to notice.}) perturbations, \ding{183} measure the model's performance on each perturbed variant, and \ding{184} calculate the maximum performance drop between consecutive perturbation levels as our sensitivity metric.

\begin{algorithm}
\scriptsize
\caption{Input Perturbation Sensitivity Evaluation}
\label{alg:pearl}
\SetAlgoLined
\KwIn{Model $M$, Prompt template $Prt$, Input $X$, Perturbation function $\sigma$}
\KwData{Perturbation rate max $pr_{max} = 5$, Answers per prompt $ans_{max} = 3$, Evaluation function $Perf$}
\KwOut{Maximum perturbation sensitivity score}

$X^* \gets [ ]$ \textcolor{gray}{\tcp*{Array of perturbed inputs}}
$Y^* \gets [ ]$ \textcolor{gray}{\tcp*{Array of model responses}}
$m \gets [ ]$ \textcolor{gray}{\tcp*{Performance differences}}

\BlankLine
\textcolor{gray}{\tcp{Generate perturbed inputs at increasing perturbation rates}}
\For{$k = 0$ \KwTo $pr_{max}$}{
    $X^*.append(\sigma(X, k))$\;
}

\BlankLine
\textcolor{gray}{\tcp{Collect LLM responses for each perturbed input}}
\For{$k = 0$ \KwTo $pr_{max}$}{
    $prompt \gets Prt(X^*[k])$\;
    $answers \gets [ ]$\;
    \For{$i = 0$ \KwTo $ans_{max}$}{
        $answers.append(M(prompt))$\;
    }
    $Y^*.append(answers)$\;
}

\BlankLine
\textcolor{gray}{\tcp{Calculate performance differences between consecutive perturbation levels}}
\For{$k = 0$ \KwTo $pr_{max}-1$}{
    $eval \gets mean(Perf(Y^*_k)) - mean(Perf(Y^*_{k+1}))$\;
    $m.append(eval)$\;
}

\KwRet{$max(m)$}
\end{algorithm}

Using this approach, we compute perturbation sensitivity scores for each benchmark and model combination. We then analyze the distribution of these scores to quantify memorization advantage at the benchmark level: a low sensitivity on a given benchmark suggests that the model has developed robust generalization capabilities in the neighborhood of the benchmark instances, maintaining consistent performance despite input variations.

Conversely, benchmarks exhibiting statistically significant higher sensitivity distributions indicate potential limitations in the model's generalization capabilities. This high sensitivity can be attributed to either:
\begin{enumerate}
    \item \textbf{Memorization effects}: The model performs well only on exact or near-exact matches to training examples but fails to maintain performance under perturbations.
    \item \textbf{Knowledge gaps}: The model lacks sufficient exposure to the underlying patterns or principles necessary to solve the perturbed variants of these problems.
\end{enumerate}

To identify statistically significant differences in sensitivity distributions, we employ the Mann-Whitney U test with Bonferroni correction for multiple comparisons, considering a significance level of $\alpha = 0.05$. This non-parametric approach avoids assumptions about the underlying distribution of sensitivity scores.
Moreover, we repeat the process 03 times while considering the average of the sensitivity for each sample considered to ensure the coherence of the results. Also, in order to mitigate the LLMs randomness, we fix the temperature at 0.3 and the $top_k$ at 0.5.

\subsection{Models Considered} \label{subsec:models}
We selected a diverse set of state-of-the-art code LLMs representing different model families, architectures, and training methodologies. This diversity enables us to investigate whether memorization advantage patterns vary systematically across different model development approaches. The models included in our analysis are:

\begin{itemize}[leftmargin=*]

    \item \textbf{DeepSeek-Coder-V2 (16B)} is an open-source family of code language models built on the DeepSeek-V2 foundation. It was pretrained on a diverse dataset comprising 60\% source code, 10\% mathematical corpus, and 30\% natural language from CommonCrawl and GitHub repositories. This balanced composition optimizes the model for a wide range of code-related tasks while maintaining strong natural language understanding capabilities~\cite{deepseek-ai_deepseek-coder-v2_2024}.
    
    \item \textbf{Qwen2.5-Coder (14B)} is a code-specialized model series built on the Qwen2.5 architecture, trained on multilingual code repositories. It implements an innovative hybrid attention mechanism specifically designed to handle both local and global code dependencies, enabling efficient processing of complex codebases with nested structures and long-range relationships~\cite{hui2024qwen2}.
    
    \item \textbf{StarCoder (15B):} is developed through an open-scientific collaboration, this open-source code model was trained on The Stack V2 corpus. The training data includes permissively licensed code, unlicensed files, and specialized datasets like APPS, CodeContest, and GSM8K to enhance mathematical reasoning and algorithmic problem-solving capabilities~\cite{lozhkov_starcoder_2024}.
    
    \item \textbf{CodeLlama (13B)} is a Meta's code-specialized model built on LLama 2, developed through a cascade of training and fine-tuning steps. This progressive approach gradually enhanced the base model's capabilities on code-related tasks while maintaining its foundation in natural language understanding, creating a versatile model for various software engineering applications~\cite{roziere_code_2024}.
    
    \item \textbf{Codestral (22B):} is an open-weight generative AI model explicitly designed for code generation tasks and trained on a diverse dataset of 80+ programming languages. Hovewer, this model is only available in the size 22B~\cite{codestral}.
    
    \item \textbf{OpenCoder (8B)} is an open and reproducible code LLM family available in two sizes (1.5B and 8B). It was pretrained using the RefineCode corpus, a reproducible dataset of 960 billion tokens across 607 programming languages, incorporating over 130 language-specific rules with customized weight assignments~\cite{Huang2024OpenCoderTO}.
    
    \item \textbf{WizardCoder (33B)} is a StarCoder variant that applies the Evol-Instruct method to evolve Code Alpaca data generated through self-instruction. The pretrained StarCoder model is then fine-tuned with this evolved data, creating a model with enhanced instruction-following capabilities for code generation tasks~\cite{luo_wizardcoder_2023}.
    
    \item \textbf{Magicoder (7B)} uses the novel OSS-Instruct methodology that leverages open-source code to create contextually relevant instruction-response pairs. This approach enables code generation that adheres to real-world best practices and community standards. The model was developed using CodeLLama-Python-7B as its base architecture~\cite{wei_magicoder_2024}.
\end{itemize}

\subsection{Benchmarks: Tasks and Datasets} \label{subsec:datasets}

We evaluate memorization advantage across three categories involving five distinct code-related tasks, using a total of 19 benchmark datasets. These tasks represent critical dimensions of software engineering that have emerged as essential benchmarks for assessing the capabilities of code LLMs, covering the full spectrum from generation to comprehension to repair. Considering these dimensions allows us to investigate whether patterns of memorization advantage vary across different types of code understanding and generation challenges.
The task categories are 
{\bf NL2Code Tasks} (Code generation), 
{\bf Code2Code Tasks} (Test generation \& Program repair) and {\bf Code2NL Tasks} (Vulnerability detection \& Code summarization).

\subsubsection{Code Generation} 
\label{subsubsec:code_gen}

Our evaluation of code generation capabilities includes six diverse benchmarks. \textbf{HumanEval}~\cite{chen2021evaluating} contains 164 hand-written programming problems with test cases, focusing on function completion tasks in Python that emphasize reasoning and algorithmic skills rather than API knowledge. \textbf{APPS}~\cite{hendrycks2021measuring}, the Assessment of Programming Problems for Students, features 10,000 problems at varying difficulty levels with input/output examples and function signatures. \textbf{MBPP}~\cite{austin2021program}, the Mostly Basic Python Programming dataset, comprises 974 Python programming tasks of simple to moderate difficulty, designed for evaluating code generation capabilities. \textbf{LBPP}~\cite{matton_leakage_2024},a collection of 161 Python programs with corresponding unit tests, specifically created to address data contamination concerns and constructed to be "fresh" (not leaked at it time of releasing in July 2024). \textbf{CodeContest}~\cite{li2022competition} provides competition-level programming problems from platforms like Codeforces, including 13,610 problems with test cases across multiple languages. Finally, \textbf{XLCost}~\cite{zhu2022xlcost}, the Cross-Lingual Code Intelligence Evaluation Suite, focuses on multilingual code generation tasks across various programming languages.

\subsubsection{Program Repair} \label{subsubsec:prog_repairs}
For program repair, we consider three established benchmarks. 
\textbf{QuixBugs}~\cite{lin2017quixbugs} is a dataset containing 40 programs derived from the Quixey Challenge, implemented in both Python and Java. Each program features a precisely identified one-line defect, accompanied by passing (when possible) and failing test cases. \textbf{Defects4J}~\cite{just2014defects4j} is a comprehensive collection of 854 reproducible bugs extracted from real-world Java projects. Each bug is isolated with a test case that triggers the failure, making it one of the most widely used benchmarks for evaluating automated program repair techniques in industrial-scale software. \textbf{ConDefects}~\cite{wu2023condefects} is an extensive benchmark comprising 1,254 faulty Java programs and 1,625 faulty Python programs, each paired with precise fault location information and corresponding repaired versions. This dataset is specifically designed to evaluate contextual program repair capabilities in a multi-language setting, offering realistic defect scenarios with ground truth fixes.

\subsubsection{Test Generation} \label{subsubsec:test_gen}
For test generation, we rely on two specialized benchmarks. \textbf{BigCodeBench}~\cite{zhuo2024bigcodebench} is a benchmark designed to evaluate LLMs with practical and challenging programming tasks that reflect real-world development scenarios. The benchmark consists of carefully curated programming problem pairs, each including an instructional prompt, canonical solution, and comprehensive unit tests, enabling rigorous assessment of models' practical coding capabilities. \textbf{TestEval}~\cite{wang2025testevalbenchmarkinglargelanguage} provides a collection of 210 Python programs sourced from online programming platforms, specifically designed to evaluate LLMs' capabilities in test case generation rather than code implementation. This benchmark assesses a model's ability to understand existing code functionality and generate appropriate test cases that verify correctness, focusing on a critical but often overlooked aspect of the software development lifecycle. We also added for this task the benchmark QuixBug which proposed for each program a unit test case associated.

\subsubsection{Vulnerability Detection} \label{subsubsec:vul_detect}
For vulnerability detection, we include six diverse benchmarks. \textbf{CVEFixes}~\cite{bhandari2021cvefixes} provides a collection of code patches that address Common Vulnerabilities and Exposures (CVEs), paired with their vulnerable counterparts. \textbf{VulDetectBench}~\cite{liu2024vuldetectbench} offers a benchmark suite for evaluating vulnerability detection systems across multiple programming languages, with annotated vulnerable code regions. \textbf{VulnPatchPairs}~\cite{risse2024uncovering} contains paired datasets of vulnerable code snippets and their corresponding security patches from real-world software. \textbf{Devign}~\cite{zhou2019devigneffectivevulnerabilityidentification} features a collection of 40 CVEs collected from 4 popular C libraries, with function-level vulnerability annotations. \textbf{ReVeal}~\cite{chakraborty2020deeplearningbasedvulnerability} includes a collection of past vulnerabilities from two open-source projects (Linux Debian Kernel and Chromium) and their associated patches. \textbf{DiverseVul}~\cite{chen2023diversevul} provides an extensive collection of 18,945 vulnerable functions across 150 Common Weakness Enumerations (CWEs) and 330,492 non-vulnerable functions, extracted from 295 projects.

\subsubsection{Code Summarization}
For code summarization, we utilize \textbf{CodeSearchNet}~\cite{husain2019codesearchnet}, a large-scale collection containing over 2 million code-docstring pairs across six programming languages (Python, Java, JavaScript, PHP, Go, and Ruby), designed for code search and summarization tasks. This benchmark enables evaluation of models' ability to generate natural language descriptions from code snippets across diverse programming paradigms and syntax. In addition, we also consider a collection of 69,708 pairs of ⟨API sequence, code, summary⟩ collected from Java projects published on Github from 2015 to 2016 and used to train the model TLCode-Sum \cite{hu2018summarizing}.

\subsection{Perturbation Methods} 
\label{subsec:perturbations}
We employ task-specific perturbation methods that 
introduce surface-level variations:

\subsubsection{Natural Language Perturbations} 
\begin{figure}[!h]
   \vspace{-0.5cm}  \centering
    \includegraphics[width=1\linewidth]{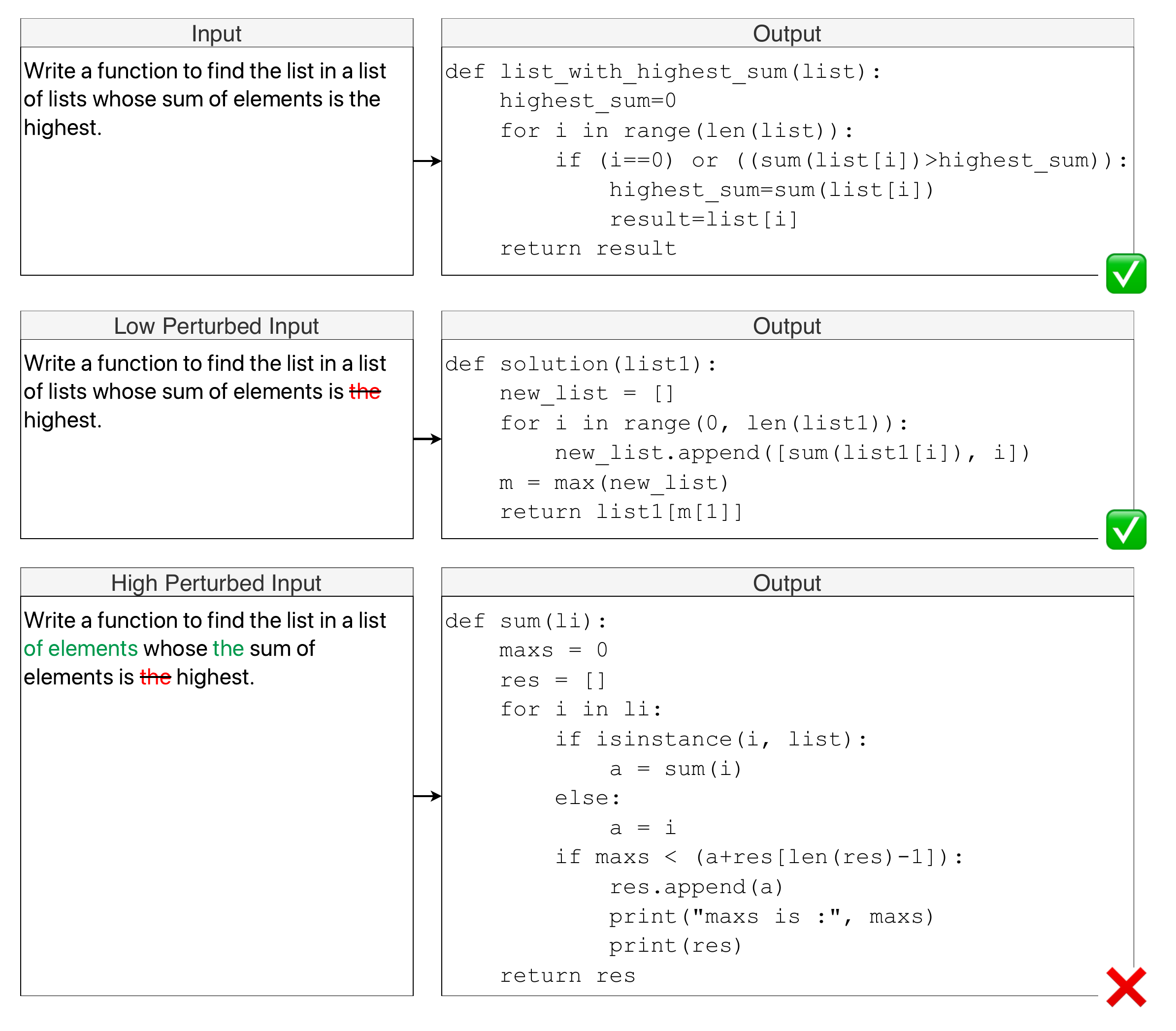}
   \vspace{-0.5cm} \caption{Examples Outputs by StarCoder2 when applying perturbations to prompt instructions from MBPP)}
    \label{fig:nl2code_sample}
    \Description{StarCoder2's performance degradation under prompt perturbations. When MBPP prompt instructions were modified, StarCoder2 maintained valid outputs for original and lightly perturbed inputs (passing test suites). However, highly perturbed prompts caused test suite failures, demonstrating the model's sensitivity to input variations.}
\end{figure}

For \textbf{NL2Code tasks}, notably code generation, with natural language prompts, we generate five progressively altered versions of each input using the BART \cite{lewis2019bart} model for controlled paraphrasing. These perturbed prompts (i.e., rephrased) are ordered by their cosine distance from the original prompt, ensuring a gradual \textit{syntactic}  drift while maintaining the core requirements and functionality described in the prompt. 
\Cref{fig:nl2code_sample} illustrates how varying intensity applied to an MBPP sample affect StarCoder2's performance in code generation. Interestingly, the outputs generated from both the original input and the low-perturbed input successfully passed all associated tests. However, the execution of the output produced from the high perturbed input fails (\textit{TypeError} on argument of $isinstance$).

\subsubsection{Code Perturbation} 
\begin{figure}[!h]
    \centering
    \includegraphics[width=1\linewidth]{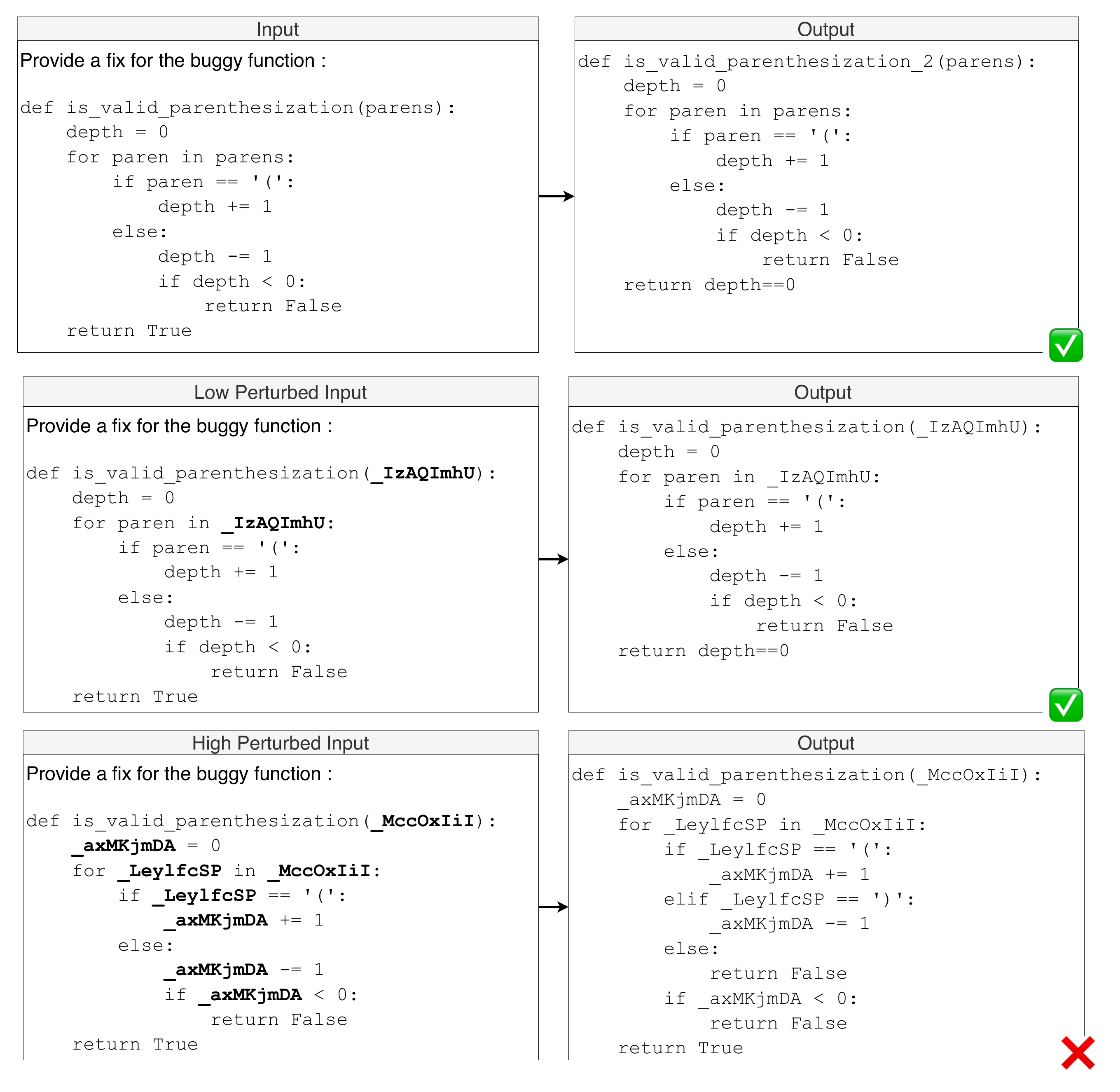}
    \caption{Example Outputs by StarCoder when apply perturbations to input prompt from QuixBugs - see supplementary file for more details about test failures}
    \label{fig:code2code_sample}
    \Description{Variable refactoring reveals StarCoder2-15B's brittleness in program repair. The model performed well on original inputs and those with minimal perturbation (single variable refactoring). However, it failed when all variables were replaced with randomly selected values (high perturbation), exposing limitations in its program understanding.}
\end{figure}

For \textbf{Code2Code or Code2NL tasks}, which take code as input, we progressively rename variables using randomly generated identifier names\footnote{We leave semantic perturbations to future work.}. The perturbation intensity increases with the number of variables renamed, creating a controlled gradient of syntactic variation. Figure \ref{fig:code2code_sample} illustrates the perturbations applied to a QuixBugs code sample at different intensity levels and the corresponding code outputs from the StarCoder2 model prompted for program repair task. We observe that the repaired code produced by the LLM for the original input and the low-perturbed input are correct (w.r.t test suite), while the ouput generated for the high-altered, but semantically equivalent, buggy versions is found to be incorrect (w.r.t test suite).

trFor both perturbation types, we maintain five distinct perturbation levels ($pr_{max} = 5$), enabling fine-grained analysis of performance degradation patterns. The perturbation functions are designed to be deterministic and reproducible, facilitating comparison across different models and experiments.

\section{Experimental Results} \label{sec:results}

This section presents our empirical findings on memorization advantage across different code-related tasks and models. For each task category, we analyze the perturbation sensitivity distributions and highlight statistically significant patterns. Lower sensitivity values indicate more robust generalization (interpolation), while higher values suggest potential memorization or knowledge gaps.

\subsection{Code Generation}

\begin{figure}[!h]
    \centering
    \includegraphics[width=0.9\linewidth]{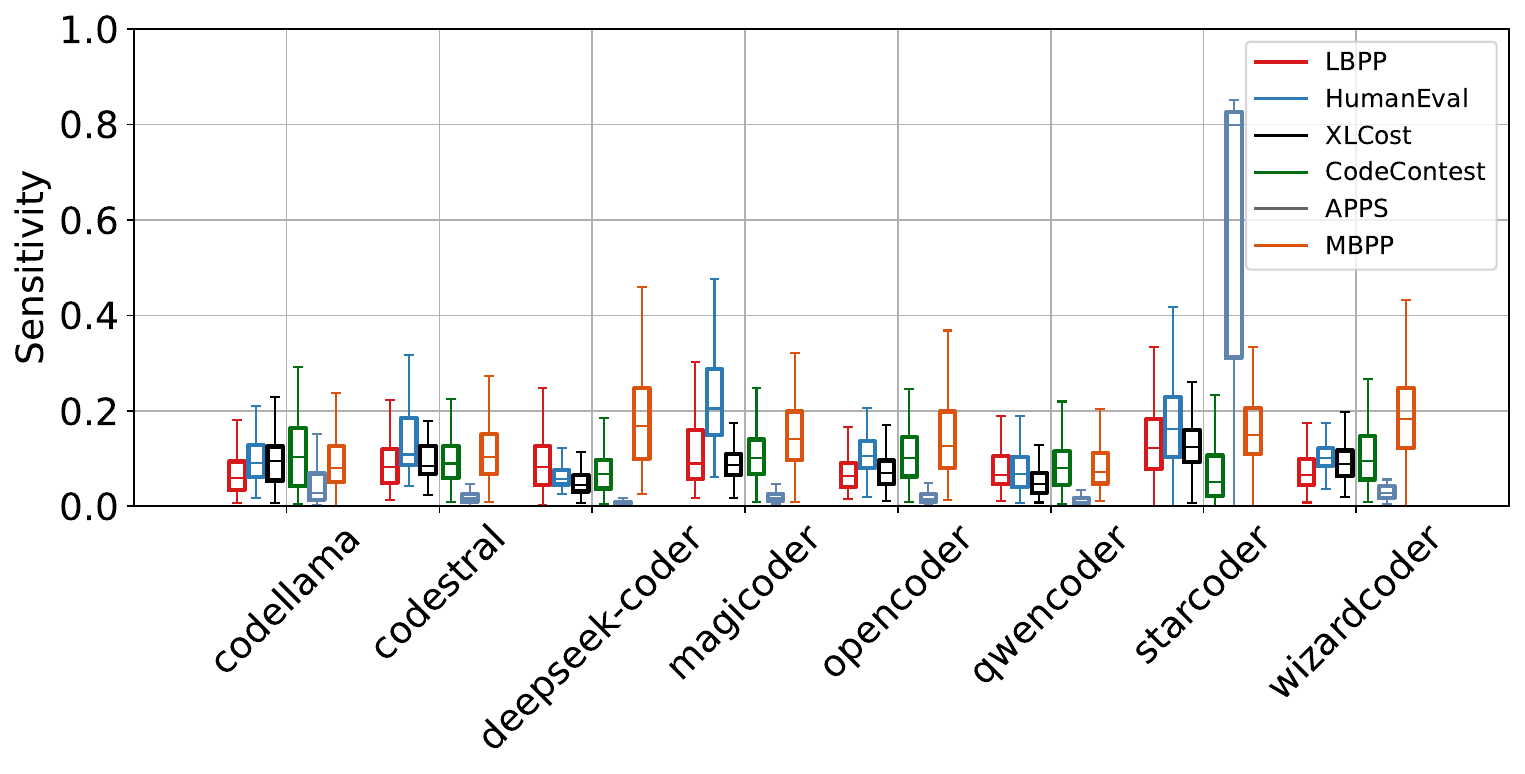}
    \vspace{-0.5cm}
    \caption{Perturbation sensitivity distributions across code generation benchmarks for all evaluated models. \normalfont \textit{Lower values indicate stronger generalization capabilities, while higher values suggest potential memorization or brittle understanding.}}
    \label{fig:code_gen_box}
    \Description{Code generation benchmark sensitivity analysis. Lower sensitivity values indicate robust generalization, while higher values suggest memorization or fragile understanding. StarCoder2-15B exhibits notably high sensitivity on the APPS benchmark compared to other models, which maintain relatively low sensitivity across benchmarks, indicating StarCoder2-15B's reliance on memorized patterns.}
\end{figure}

Figure~\ref{fig:code_gen_box} illustrates the sensitivity distributions across code generation benchmarks. The majority of models exhibit relatively low sensitivity values ($<0.4$) across most code generation benchmarks, suggesting robust interpolation capabilities. QwenCoder and CodeLlama consistently demonstrate the lowest sensitivity scores, indicating superior generalization capabilities for fundamental coding tasks.

Among the models evaluated, StarCoder exhibits a markedly different pattern. It demonstrates significantly elevated sensitivity on the APPS benchmark ($\sim0.8$, $p < 0.01$ using Mann-Whitney U test with Bonferroni correction), substantially deviating from both its performance on other benchmarks and from other models' performance on APPS. This pronounced elevation suggests potential memorization behavior specifically for this benchmark, which may indicate training data contamination.

When comparing across benchmarks, we find that the MBPP benchmark consistently elicits elevated sensitivity values across the model spectrum compared to other code generation tasks. This suggests that MBPP's problems may require more specific knowledge patterns that are less amenable to generalization, despite their ostensibly basic nature.

\find{{\bf  Key Insights \ding{42} Code Generation } 
    \begin{itemize}[leftmargin=*]
    \item Most models demonstrate strong generalization capabilities for basic code generation tasks
    \item StarCoder shows significant evidence of potential memorization specifically on the APPS benchmark
    \item MBPP presents unique generalization challenges across all evaluated models, suggesting its problems may require more specific knowledge patterns
    \item QwenCoder and CodeLlama exhibit the most robust generalization profiles for code generation
\end{itemize}
}

\subsection{Test Generation}

\begin{figure}[!h]
    \centering
    \includegraphics[width=0.9\linewidth]{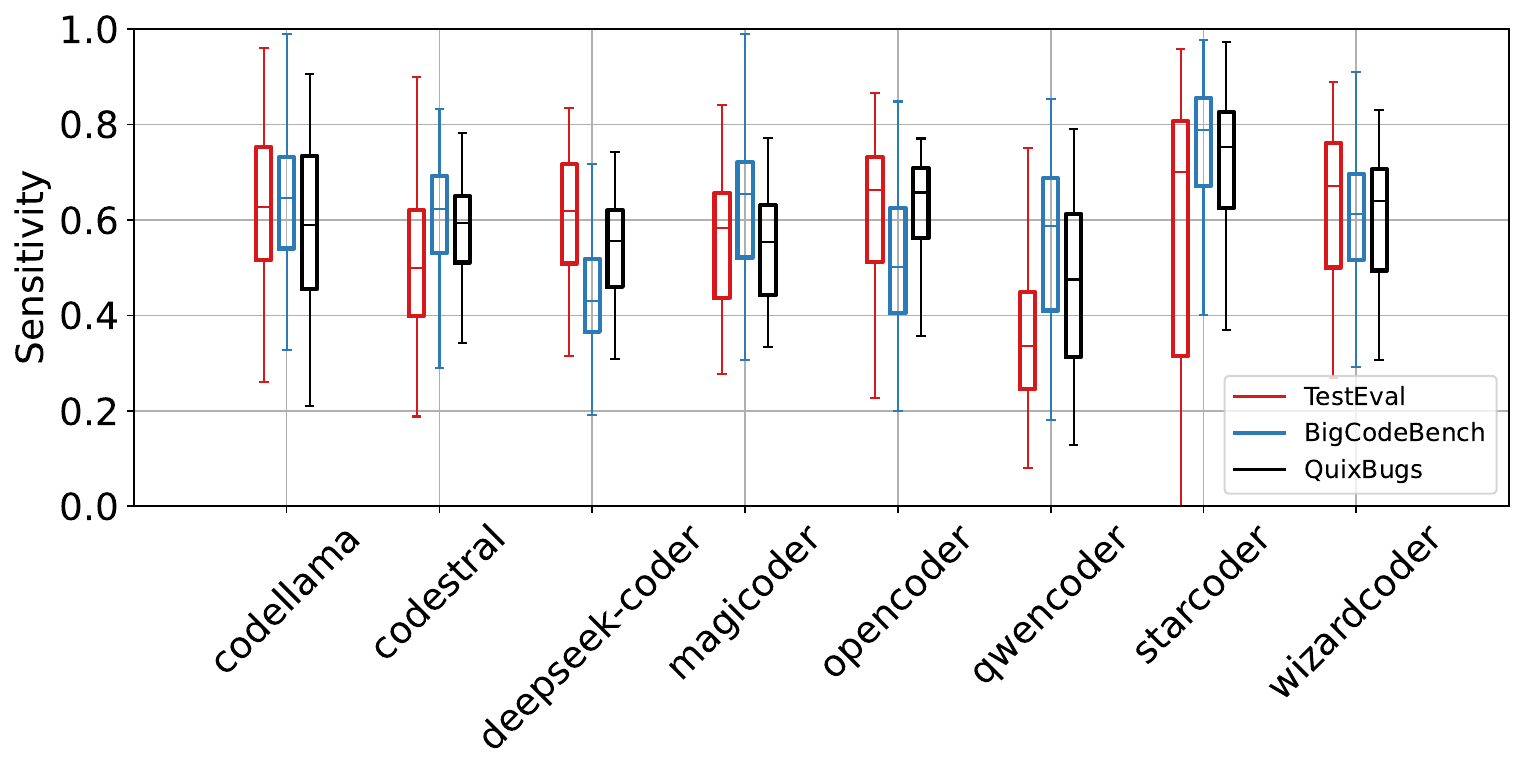}
    \vspace{-0.5cm}
    \caption{Perturbation sensitivity distributions across test generation benchmarks. \normalfont \textit{Higher values indicate greater sensitivity to input perturbations, suggesting potential challenges in generalizing test generation capabilities}.}
    \label{fig:test_gen_box}
    \Description{Test generation benchmark sensitivity analysis. Lower sensitivity values indicate robust generalization, while higher values suggest memorization or fragile understanding. All evaluated models demonstrate high sensitivity patterns across test generation benchmarks, indicating widespread challenges in generalizing beyond training data for this task.}
\end{figure}

Figure~\ref{fig:test_gen_box} presents sensitivity distributions for test generation benchmarks. Test generation requires deeper code comprehension capabilities than basic code generation. All evaluated models demonstrate substantially higher sensitivity values for test generation tasks ($0.4$-$0.7$ median range) compared to basic code generation tasks ($0.2$-$0.4$ median range). This statistically significant difference ($p < 0.001$, paired t-test) suggests that tasks requiring deeper code understanding are more susceptible to perturbation effects, reflecting the increased complexity of generating tests versus implementing functionality.

The consistency across benchmarks reveals that TestEval, QuixBugs, and BigCodeBench exhibit remarkably similar sensitivity patterns across models (Pearson correlation $r > 0.85$ between benchmark sensitivities). This high correlation suggests these benchmarks evaluate fundamental test generation capabilities that transcend specific model architectures or training methodologies, revealing inherent generalization boundaries for this task. StarCoder consistently registers sensitivity values at the upper bound of the distribution, reinforcing its tendency toward higher sensitivity observed in other task categories.

\find{{\bf  Key Insights \ding{42} Test Generation } 
    \begin{itemize}[leftmargin=*]
        \item Test generation tasks consistently show higher perturbation sensitivity than code generation across all models, which suggests that  test generation represents a more challenging generalization task for current code LLMs
        \item All three test generation benchmarks exhibit similar sensitivity patterns, suggesting similar fundamental test targets/patterns
    \end{itemize}
}

\subsection{Program Repair}

\begin{figure}[!h]
    \centering
    \includegraphics[width=0.9\linewidth]{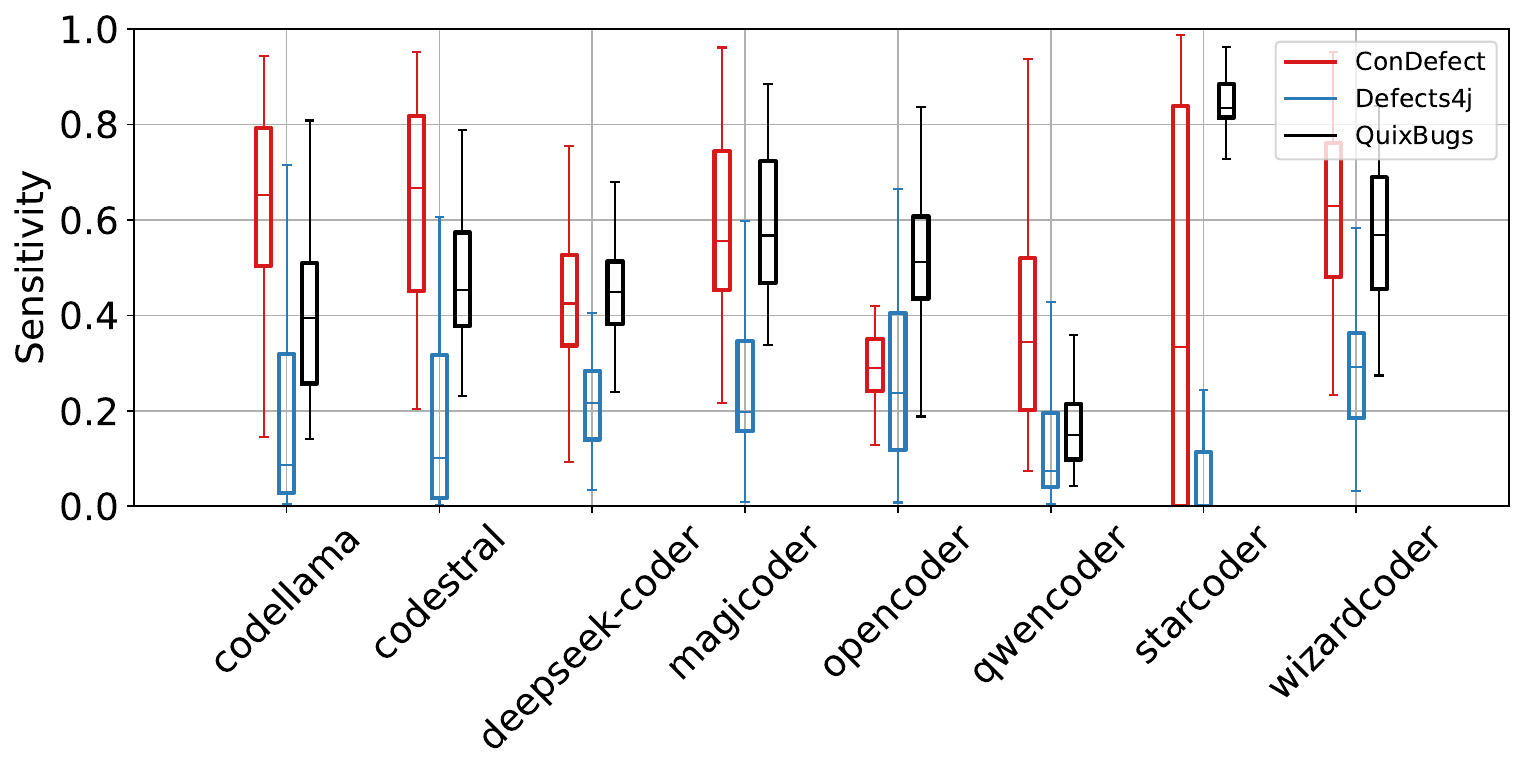}
    \vspace{-0.5cm}
    \caption{Perturbation sensitivity distributions for program repair benchmarks. \normalfont \textit{The varying sensitivity patterns across benchmarks suggest different levels of generalization challenges for bug detection and repair tasks}.}
    \label{fig:prog_repair_box}
    \Description{Program repair benchmark sensitivity analysis. Lower sensitivity values indicate robust generalization, while higher values suggest memorization or fragile understanding. All models show high sensitivity across program repair benchmarks, with Defects4J being the notable exception where models demonstrate better generalization.}
\end{figure}

Figure~\ref{fig:prog_repair_box} presents sensitivity distributions for program repair benchmarks. Defects4J demonstrates notably lower sensitivity ($0.2$-$0.4$) compared to other program repair benchmarks ($0.5$-$0.8$), representing a statistically significant difference ($p < 0.01$, Mann-Whitney U test). This unexpected robustness challenges prevailing concerns about data leakage in this widely-used benchmark, suggesting models may have developed genuine generalization capabilities for the types of bugs represented in Defects4J rather than merely memorizing specific instances.

Benchmark-specific patterns reveal varying challenges. ConDefect consistently elicits the highest sensitivity values across most models (median $\sim0.7$), indicating that repair tasks requiring broader contextual understanding present particular generalization challenges. This aligns with the benchmark's design focus on bugs that span multiple functions or require module-level comprehension, suggesting current models' generalization capabilities may be more limited for context-dependent repairs.

StarCoder exhibits exceptionally high sensitivity on bug detection tasks, particularly evident with the ConDefect benchmark, while QwenCoder lacks measurable results for certain benchmarks in this category, potentially indicating limitations in its training data coverage for program repair tasks.

\find{{\bf  Key Insights \ding{42} Program Repair } 
    \begin{itemize}[leftmargin=*]
    \item Defects4J shows surprisingly low sensitivity, challenging literature concerns about data leakage in this widely-used benchmark
    \item Context-dependent bugs (ConDefect) present the greatest generalization challenge across all models
    \item The variability in sensitivity across repair benchmarks indicates differing levels of generalization difficulty based on bug context and complexity
    \item Models demonstrate varying generalization capabilities for different types of bugs, suggesting specialized knowledge
\end{itemize}
}

\subsection{Vulnerability Detection}

\begin{figure}[!h]
    \centering
    \includegraphics[width=0.9\linewidth]{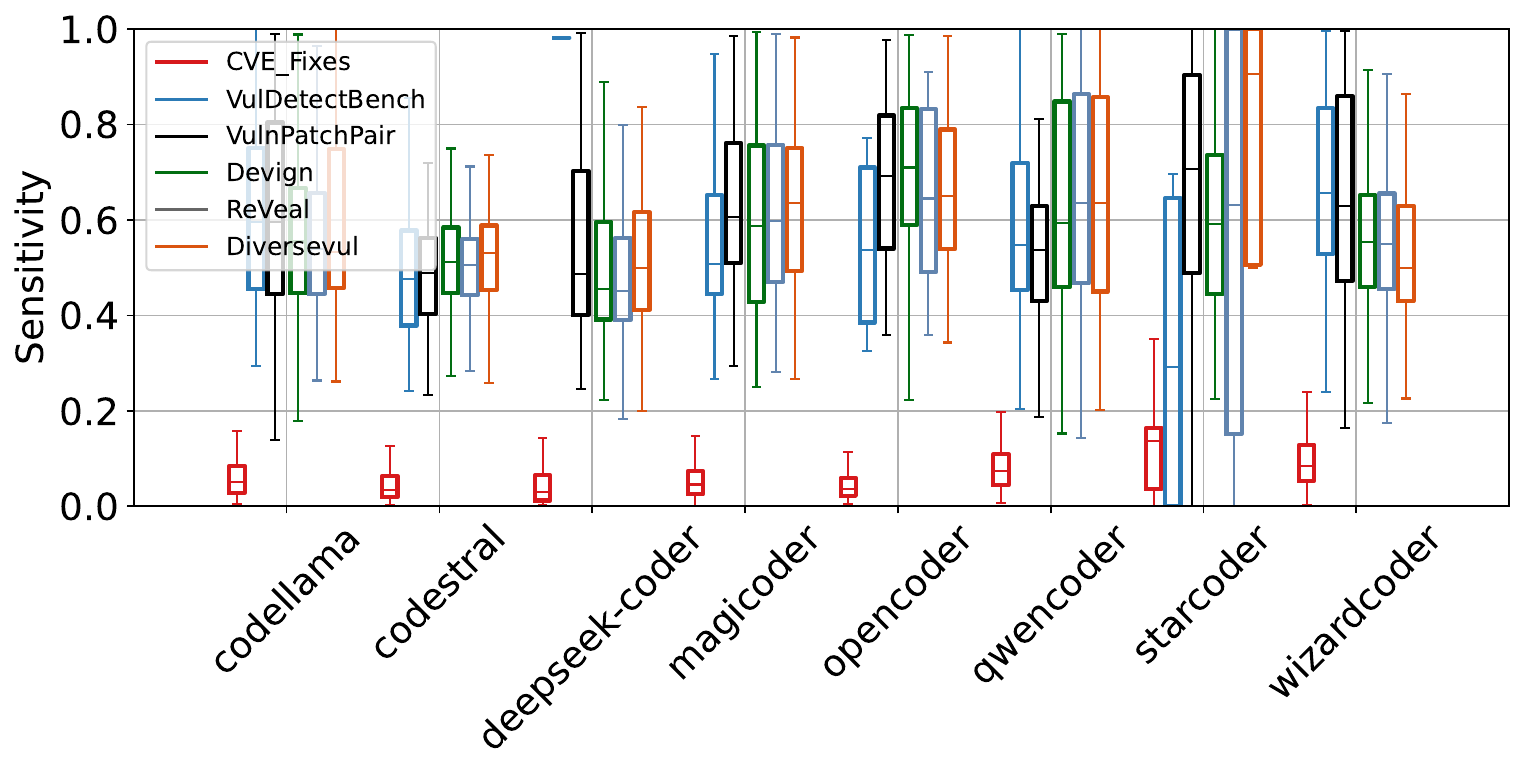}
    \vspace{-0.5cm}
    \caption{Perturbation sensitivity distributions across vulnerability detection benchmarks. \normalfont \textit{Note the exceptionally low sensitivity for CVEFixes across all models, contrasting with moderate to high sensitivity for other security benchmarks.}}
    \label{fig:vul_detect_box}
    \Description{Vulnerability detection benchmark sensitivity analysis. Lower sensitivity values indicate robust generalization, while higher values suggest memorization or fragile understanding. Models exhibit high sensitivity across vulnerability detection benchmarks, except for CVEFixes where they show improved generalization capabilities.}
\end{figure}

Figure~\ref{fig:vul_detect_box} illustrates sensitivity distributions across security-oriented benchmarks. The security domain presents the widest range of sensitivity distributions among all task categories (ranging from $<0.1$ to $>0.8$), suggesting pronounced variability in generalization capabilities for security-related tasks. This variability likely reflects the diverse nature of security vulnerabilities and the different approaches needed to identify them.

Particularly noteworthy is the exceptional robustness observed in one benchmark: CVEFixes demonstrates exceptionally low sensitivity (consistently below $0.1$) across all evaluated models, representing a statistically significant deviation ($p < 0.001$, Kruskal-Wallis test) from other security benchmarks. This unexpected robustness challenges previous assumptions about potential data leakage in this benchmark and suggests models have developed strong, generalized understanding of common vulnerability patterns rather than memorizing specific CVE instances.

The remaining security benchmarks (VulDetectBench, VulnPatchPair, Devign, ReVeal, and DiverseVul) present similar sensitivity distributions with moderate to high values ($0.4$-$0.8$) across models. This consistency suggests these benchmarks evaluate similar underlying capabilities related to vulnerability identification, with the higher sensitivity values indicating ongoing challenges in generalizing security knowledge.

\find{{\bf  Key Insights \ding{42} Vulnerabilty Detection } 
\begin{itemize}[leftmargin=*]
    \item Security benchmarks show the widest variance in sensitivity among all task categories
    \item CVEFixes exhibits remarkably low sensitivity across all models, challenging assumptions about data leakage
    \item Most security benchmarks show moderate to high sensitivity, indicating generalization challenges for security tasks
    \item The exceptional performance on CVEFixes suggests models may have successfully generalized common vulnerability patterns, rather than memorizing specific instances
\end{itemize}
}

\subsection{Code Summarization}
\begin{figure}[!h]
   \centering
   \includegraphics[width=0.9\linewidth]{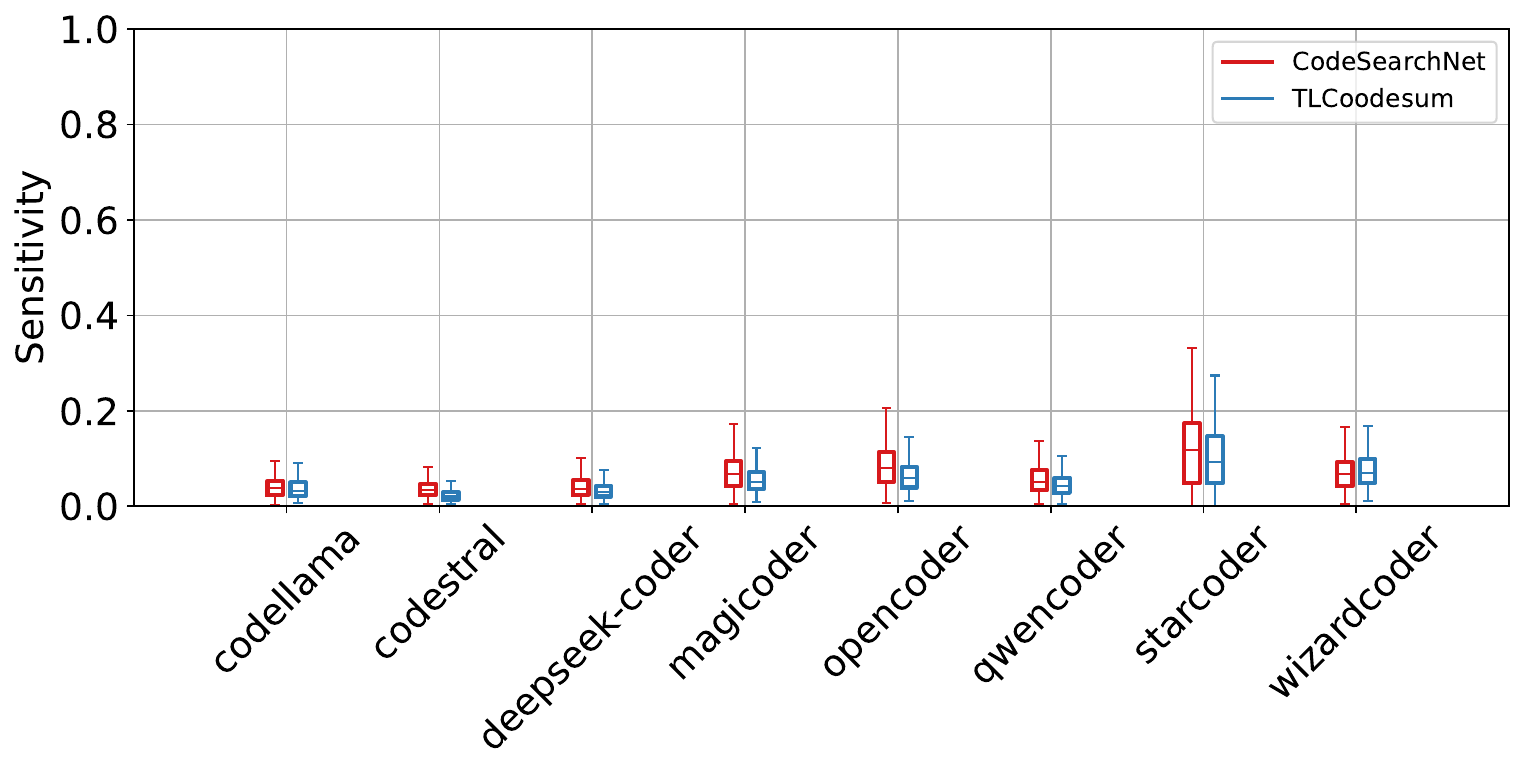}
    \vspace{-0.5cm}
\caption{Perturbation sensitivity distributions for code summarization benchmarks. \normalfont \textit{The consistently low sensitivity values across all models suggest robust generalization capabilities for this task category}.}
   \label{fig:code_sum_box}
   \Description{Code summarization benchmark sensitivity analysis. Lower sensitivity values indicate robust generalization, while higher values suggest memorization or fragile understanding. All evaluated models demonstrate low sensitivity patterns across code summarization benchmarks, indicating strong generalization capabilities for this task type.}
\end{figure}

Figure~\ref{fig:code_sum_box} illustrates the sensitivity distribution in code summarization benchmarks. All evaluated models demonstrate remarkably low sensitivity values for code summarization tasks (median values $<0.3$), suggesting robust interpolation capabilities across the model spectrum. This indicates that models have developed generalizable understanding of the relationship between code structure and natural language descriptions, making this task less susceptible to perturbation effects than other code-related tasks.

The consistently low sensitivity values observed across different model architectures suggest that code summarization may be a task where current code LLMs have developed particularly strong generalization capabilities, possibly due to the abundance of code-documentation pairs in pre-training datasets and the more standardized nature of the task compared to open-ended code generation or complex program repair.

\find{{\bf  Key Insights \ding{42} Code Summarization } 
\begin{itemize}[leftmargin=*]
    \item All models demonstrate strong generalization capabilities for code summarization tasks
    \item The consistently low sensitivity across different model architectures suggests code summarization may be a task where current LLMs excel
    \item Potential explanations include abundant code-documentation pairs in pre-training data and the relatively standardized nature of the task
    \item Code summarization represents the task with the most robust generalization among all evaluated categories
\end{itemize}
}
\section{Discussion} \label{sec:discussion}

Our analysis of memorization advantage across code LLMs reveals important insights about model generalization capabilities, benchmark reliability, and evaluation methodologies. In this section, we discuss the broader implications of our findings, their limitations, and directions for future research.

\subsection{Key Findings and Their Implications}

Our perturbation sensitivity analysis revealed several notable patterns with important implications for code LLM development and evaluation:

\subsubsection{Task-Dependent Generalization Capabilities}

Our results show that code LLMs exhibit varying degrees of generalization capabilities across different tasks. 
This task-dependent pattern suggests that different capabilities may require different evaluation approaches. For tasks where models have developed strong generalization (e.g., code summarization), standard benchmarks may provide reliable assessments. However, for tasks with higher sensitivity (e.g., test generation), researchers should employ more rigorous evaluation methods that account for potential memorization effects.

\subsubsection{Benchmark Contamination vs. Genuine Generalization}

One of our most surprising findings was the exceptionally low sensitivity observed for certain benchmarks that have been previously suspected of data contamination, particularly CVEFixes and Defects4J. This finding challenges the common assumption that strong performance on widely available benchmarks necessarily indicates memorization or data leakage. It suggests that some benchmark patterns may be easier to generalize, perhaps due to their similarity to common programming patterns or because they represent fundamental concepts that are well represented in diverse training data.  It is possible that very high duplication of a particular datum may train models to tolerate more noise around that datum, achieving a sort of locally constrained generalization to the immediate neighborhood of a memorized datum, while still not achieving actual generalization, but instead just pushing the task performance sensitivity farther out from the datum.  This may be what we are seeing with CVEFixes and Defects4J.

\subsubsection{Model-Specific Patterns}

The significant variations in sensitivity patterns across models with comparable parameter counts (e.g., StarCoder vs. CodeLlama) indicate that architectural design choices and training methodologies substantially impact generalization capabilities. This finding has important implications for model development, confirming recent findings (cf. DeepSeek) that simply scaling up model size or training on more data may be less effective than thoughtful architectural design and training methodology. The consistent generalization advantages of instruction-tuned models may suggest that alignment techniques may improve not only safety but also fundamental generalization capabilities.

\subsection{Mitigating Memorization Advantage\newline in Evaluation}

Our findings highlight the need for more robust evaluation frameworks that account for potential memorization advantage effects. We propose several approaches to mitigate these issues:

\subsubsection{Dynamic Benchmarks}

Static benchmarks present inherent risks of contamination as they become widely used and potentially included in training datasets. Dynamic benchmarks that are regularly updated with new examples offer a promising alternative. By continuously refreshing benchmark content, the probability of data leakage is minimized, and evaluations better reflect true model capabilities rather than memorization effects.

Implementation of dynamic benchmarks could follow several strategies:

\begin{itemize}[leftmargin=*]
    \item \textbf{Time-stamped versioning}: Creating benchmark versions with clear temporal boundaries allows researchers to select versions predating a model's training cutoff.
    \item \textbf{Procedural generation}: For certain tasks, programmatically generating new benchmark examples with controlled complexity and characteristics can ensure evaluation on unseen examples.
    \item \textbf{Human-in-the-loop curation}: Incorporating expert input to continuously develop new benchmark examples that target specific capabilities while avoiding patterns from existing benchmarks.
\end{itemize}

\subsubsection{Perturbation-Based Evaluation}
Our research demonstrates the value of perturbation-based evaluation approaches in assessing memorization advantage. We recommend incorporating perturbation sensitivity analysis as a standard component of code LLM evaluation frameworks. This approach provides deeper insights than performance metrics alone and can help identify potential memorization effects.

Specifically, we recommend:


\noindent
\wcircle{1}
Reporting perturbation sensitivity alongside standard performance metrics.

\noindent
\wcircle{2}
Employing diverse perturbation types that preserve semantic meaning.

\noindent
\wcircle{3}
Setting benchmark-specific thresholds for acceptable sensitivity levels.

\noindent
\wcircle{4}
Comparing sensitivity distributions across models on the same benchmarks.

\subsubsection{Cross-Domain Evaluation}

Our results revealed that models often exhibit different sensitivity patterns across task domains. This suggests that comprehensive evaluation should include diverse tasks spanning multiple domains to provide a more complete picture of a model's generalization capabilities.

We recommend evaluation frameworks that systematically assess models across the full spectrum of code-related tasks, from generation to comprehension to repair. This approach helps identify domain-specific strengths and weaknesses while providing a more nuanced understanding of overall generalization capabilities.

\subsection{Theoretical Implications}

Our findings contribute to the broader theoretical understanding of memorization and generalization in large language models:

\subsubsection{Memorization as a Spectrum}

The varied sensitivity patterns observed across tasks and benchmarks suggest that memorization versus generalization is not a binary distinction but rather a spectrum. Models may partially memorize certain patterns while developing genuine generalization capabilities for others, and the boundary between these processes is often blurry.

This perspective aligns with recent theoretical work suggesting that memorization may be a necessary precursor to generalization~\cite{feldman2020neural, carlini_extracting_nodate}. Our results provide empirical support for this view in the context of code LLMs, showing that even models with strong generalization capabilities exhibit varying degrees of sensitivity across different tasks and benchmarks.

\subsubsection{Task-Specific Generalization Boundaries}

The consistent sensitivity patterns observed within task categories suggest that different tasks may have inherent generalization boundaries—thresholds beyond which current model architectures struggle to develop robust, generalizable capabilities. These boundaries appear to correlate with task complexity and the degree to which tasks require deep semantic understanding versus pattern recognition.

This finding has implications for model architecture design, suggesting that specialized architectures or training methodologies may be necessary to overcome generalization boundaries for particularly challenging tasks. It also suggests that evaluation frameworks should account for these task-specific boundaries when interpreting performance metrics.

\subsection{Limitations and Threats to Validity}

While our study provides valuable insights into memorization advantage in code LLMs, several limitations and threats to validity should be acknowledged:

\noindent\textbf{{\em Unknown Training Data.}}
A fundamental limitation of our study is the lack of complete knowledge about the specific datasets used to train the evaluated models. Without definitive information about training data composition, we cannot make absolute claims about data leakage. Our perturbation sensitivity analysis provides strong evidence but cannot definitively prove memorization versus generalization in all cases.

\noindent\textbf{{\em Perturbation Method Limitations.}}
Our perturbation methods focus on specific types of input variations that preserve semantic meaning. While these methods are effective for detecting certain types of memorization, they may not capture all possible forms of memorization or generalization. Different perturbation types might reveal different sensitivity patterns, and certain models might be more robust to particular perturbation types.
We use a single code perturbation: alpha-renaming of identifiers to nonces.  
Cao et al. showed that LLMs can tolerate substantial noise in their inputs~\cite{cao-etal-2023-unnatural}: it may be that our code perturbation does not, for some memorized instances, push the model beyond its noise tolerance to expose a sudden performance drop.

\noindent\textbf{{\em Benchmark Selection.}}
While we included a diverse set of benchmarks spanning multiple task categories, our selection is not exhaustive. Different benchmarks within the same task category might reveal different sensitivity patterns. Additionally, newly emerging tasks and benchmarks might present different generalization challenges not captured in our analysis.

\noindent\textbf{{\em Model Version Specificity.}}
Our findings are specific to the particular versions of the models evaluated. Newer versions of these models, or models with different parameter counts or training methodologies, might exhibit different sensitivity patterns. The field of code LLMs is rapidly evolving, and our results represent a snapshot of current capabilities.

\subsection{Future Research Directions}

Based on our findings and limitations, we identify several promising directions for future research:

\noindent\textbf{{\em Architectural Innovations for Task-Specific Generalization.}}
Our results highlight specific tasks and benchmarks where current models struggle to develop robust generalization capabilities. Future research could focus on developing specialized architectural components or training methodologies targeting these challenging domains, particularly test generation and context-dependent program repair.

\noindent\textbf{{\em Expanded Perturbation Methodologies.}}
Developing more diverse and sophisticated perturbation methodologies would provide deeper insights into model generalization capabilities. Future work could explore other perturbations, including those that perturb semantics.

\noindent\textbf{{\em Longitudinal Studies of Model Evolution.}}
Tracking changes in perturbation sensitivity across different versions of the same model family would provide valuable insights into how generalization capabilities evolve with advances in architecture, training data, and alignment techniques. Such longitudinal studies could help identify the innovations that most effectively improve generalization.

\noindent\textbf{{\em Human-Model Comparative Studies.}}
Comparing the perturbation sensitivity patterns of code LLMs with those of human programmers would provide interesting insights into the differences between machine and human generalization. Such studies could help identify areas where models have developed human-like robustness versus areas where they rely on more brittle pattern matching.

\section{Conclusion}
\label{sec:conclusion}

Our large-scale investigation into memorization advantage in code LLMs provides important insights for both researchers and practitioners. By quantifying perturbation sensitivity across diverse tasks and benchmarks, we have revealed task-specific, model-specific, and benchmark-specific patterns that challenge existing assumptions about memorization and generalization.

These findings highlight the need for more nuanced evaluation frameworks that account for potential memorization effects while recognizing that strong performance on widely used benchmarks does not necessarily indicate data leakage. As code LLMs continue to advance and see broader adoption in software development workflows, understanding these generalization boundaries becomes increasingly important for developing reliable, robust systems.

The significant variations in sensitivity patterns across models with comparable parameter counts suggest that architectural design choices and training methodologies substantially impact generalization capabilities—a finding that has important implications for future model development. By focusing not just on absolute performance metrics but also on perturbation robustness, the field can develop more genuinely capable models that generalize effectively across the full spectrum of code-related tasks.
Moreover, 

\noindent{\bf Open Science.}
\label{sec:data_availability}
To promote transparency and facilitate reproducibility, we make our artifacts available to the community at: 
\begin{center}
\url{https://github.com/Berickal/CodeLLM_Memo.git}
\end{center}
The repository includes the experiment scripts and the results. In addition, A supplementary file containing complementary analysis has been included to provide further details about this work.

\section{Acknowledgements}

This work was supported by (1) the Luxembourg Ministry of Foreign and European Affairs through
their Digital4Development (D4D) portfolio under the LuxWAyS project and (2) the European Research
Council (ERC) under the European Union’s Horizon 2020 research and innovation program
(Project NATURAL - Grant agreement N° 949014).
\balance
\bibliographystyle{ACM-Reference-Format}
\bibliography{main}

\end{document}